%% file: 2017-10-01_LYates_uBooNE-DL-LEE_Proceedings.tex
\documentclass[11pt]{article}
\usepackage{amsmath,booktabs,graphicx,url}

\input{econfmacros.tex}

\def\Title#1{\begin{center} {\Large {\bf #1} } \end{center}}

\begin{document}

\Title{MicroBooNE Investigation of Low-Energy Excess Using Deep Learning Algorithms}

\bigskip\bigskip

\begin{raggedright}  
Lauren E.\ Yates\\
\textit{Department of Physics\\
Massachusetts Institute of Technology\\
Cambridge, Massachusetts, USA}\\
\bigskip
On behalf of the MicroBooNE Collaboration\\
\bigskip\smallskip
Talk presented at the APS Division of Particles and Fields Meeting (DPF 2017), July 31--August 4, 2017, Fermilab. C170731.
\bigskip\bigskip
\end{raggedright}

\begin{abstract}
MicroBooNE is a neutrino experiment based at Fermilab which consists of a liquid argon time-projection chamber in the Booster Neutrino Beam (BNB). The experiment aims to investigate the excess of electron-neutrino-like events seen by the MiniBooNE experiment, also located in the BNB, which is potential evidence for new non-Standard Model physics such as sterile neutrinos. I discuss the status of a search for low-energy electron-neutrino interactions within the MicroBooNE detector. This analysis features a hybrid approach of traditional reconstruction methods along with the use of convolutional neural networks (CNNs), a type of deep learning algorithm highly adept at pattern recognition. I describe the identification of events and the ways in which the CNNs are used. I also outline the ways that we are addressing issues related to applying CNNs, which are trained on simulated data, to data from the detector.
\end{abstract}

\section{Introduction}

\subsection{MiniBooNE Low-Energy Excess}
\begin{figure}
\begin{center}
\includegraphics[width=0.85\textwidth]{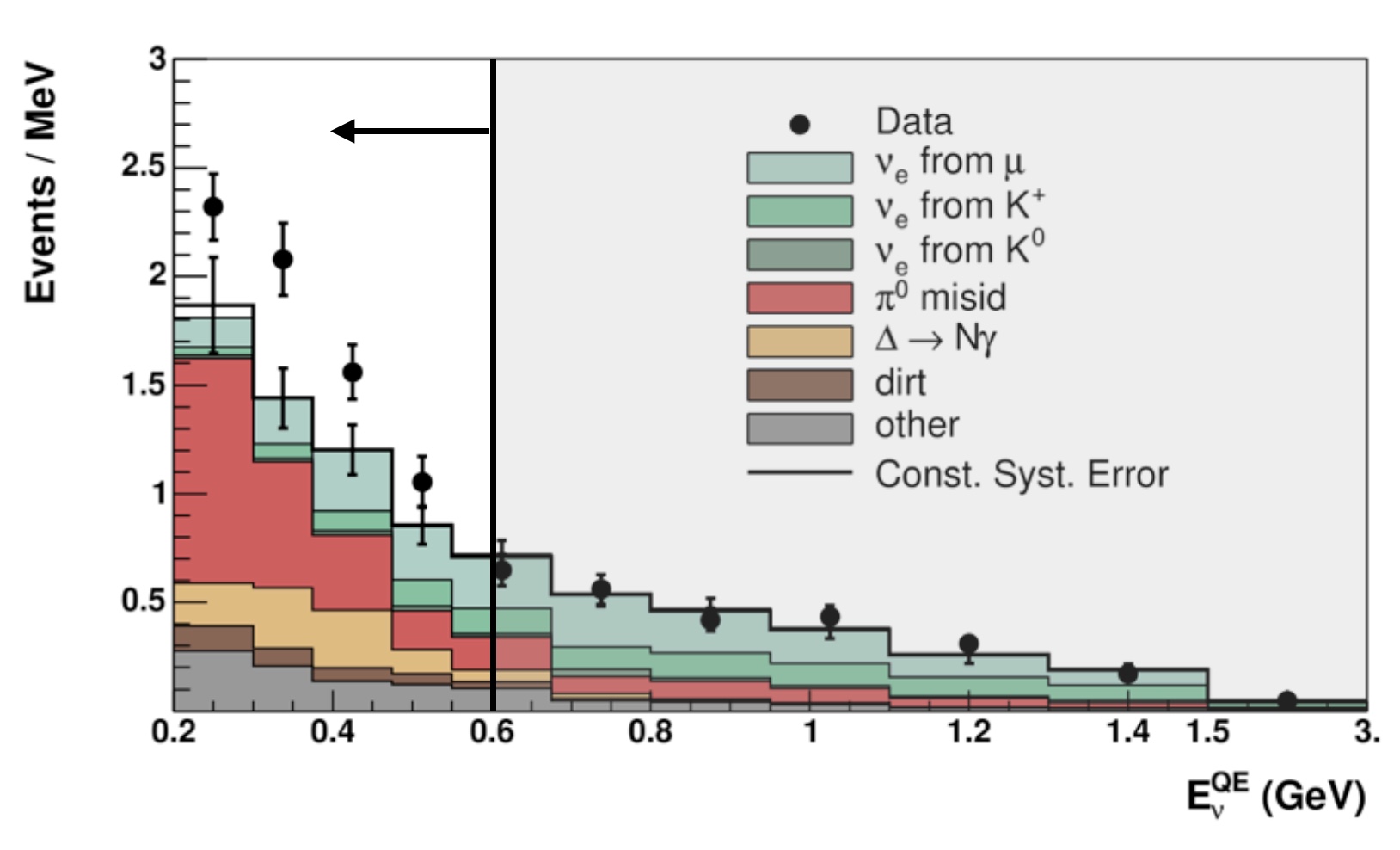}
\caption{Plot of the number of electron-neutrino-like events observed by the MiniBooNE experiment as a function of reconstructed neutrino energy~\cite{miniboone_excess}. The excess in the data compared to the stacked backgrounds in the 200 to 600~MeV energy range is highlighted.}
\label{fig:miniboone_plot}
\end{center}
\end{figure}

The primary motivation for this work is the excess of low-energy electron-neutrino-like events observed by the MiniBooNE experiment~\cite{miniboone_excess}.
MiniBooNE is a mineral oil Cherenkov detector in the Booster Neutrino Beam (BNB) at Fermilab. The BNB is a primarily $\nu_\mu$ beam with a small intrinsic $\nu_e$ component and has an average neutrino energy of about 800~MeV~\cite{bnb_flux}.
As shown in Figure~\ref{fig:miniboone_plot}, the MiniBooNE experiment saw an approximately 3$\sigma$ excess of electron-neutrino-like events with a reconstructed neutrino energy of 200 to 600~MeV.
However, this excess is in tension with global fits to a $3+1$ sterile neutrino model~\cite{cacs}.

MicroBooNE is well-positioned to investigate this anomaly.
The MicroBooNE detector, a liquid argon time projection chamber (LArTPC), is in the same beam and is at a similar baseline.
However, the LArTPC detector technology provides significantly better ability to distinguish photons from electrons and to reject other backgrounds. This will reduce the level of the backgrounds in a MicroBooNE electron neutrino appearance plot comparable to Figure ~\ref{fig:miniboone_plot}, especially the contributions that are related to photon misidentification in a Cherenkov detector (shown in red and yellow in the stacked background histogram).

\subsection{MicroBooNE}
\label{sec:uboone}
MicroBooNE is designed for precision neutrino physics measurements~\cite{uboone_detector}. The LArTPC detector combines high spatial resolution with calorimetry for excellent particle identification.

\begin{figure}
\begin{center}
\includegraphics[width=0.85\textwidth]{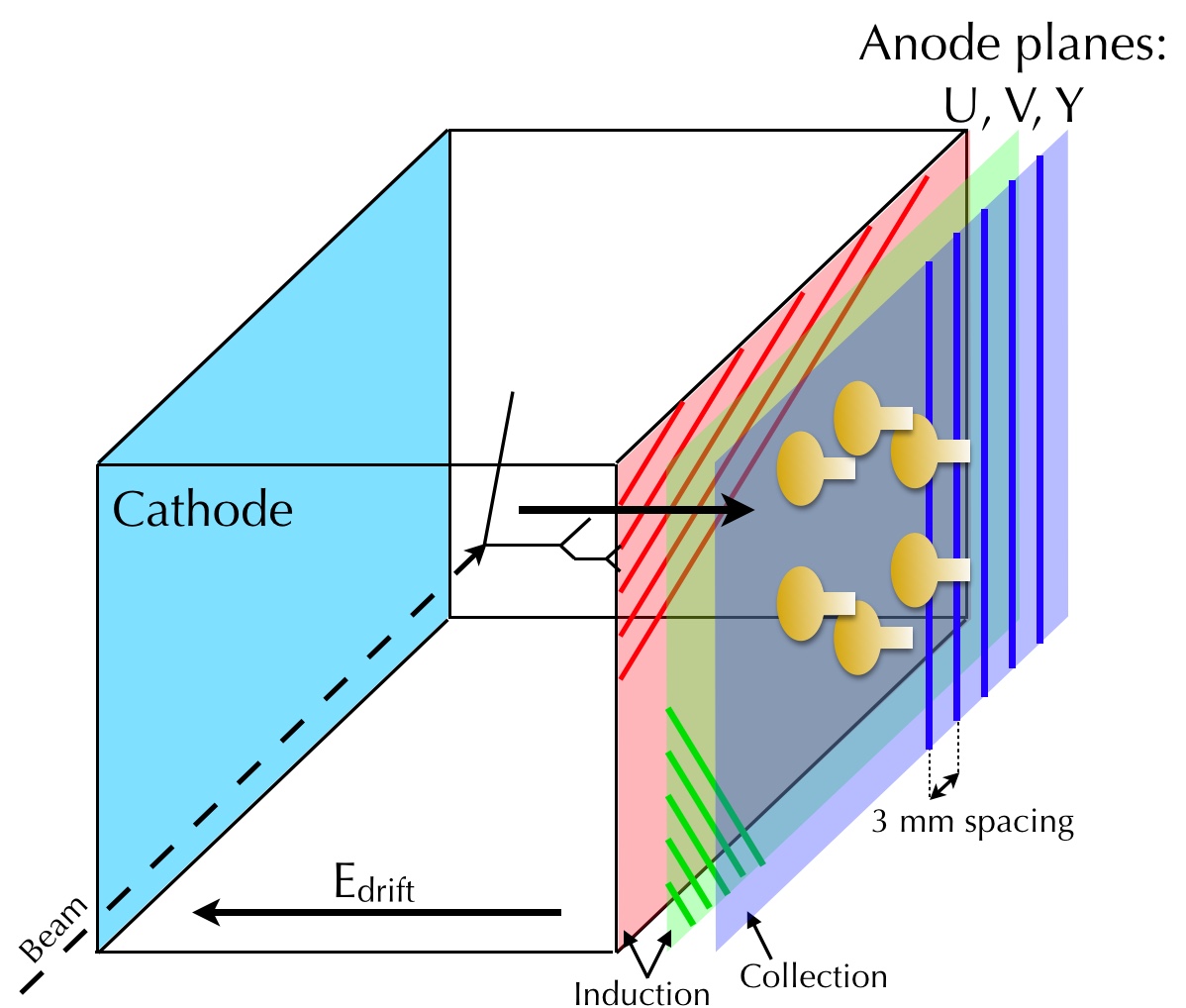}
\caption{Sketch of the MicroBooNE detector described in Section~\ref{sec:uboone}.}
\label{fig:detector}
\end{center}
\end{figure}

When a neutrino interaction occurs in the detector, it can produce charged particles. These charged particles produce two types of information that are collected by the MicroBooNE detector: scintillation light and ionization electrons. A sketch of the detector is shown in Figure~\ref{fig:detector}.
The TPC is 10.4~m long in the beam direction, 2.5~m wide in the drift direction, and 2.3~m tall. It is located inside a cylindrical cryostat and is immersed in liquid argon.
The mass of liquid argon in the TPC is about 90~tonnes.

The scintillation light is measured by an array of 32 8-inch photomultiplier tubes (PMTs).
Each PMT has an acrylic plate in front of it, which is coated with the organic wavelength-shifter tetraphenyl butadiene (TPB).
The TPB shifts the argon scintillation light down to a wavelength at which PMTs are sensitive.

The ionization electrons are drifted by an electric field and measured by three wire planes. 
We operate with an electric field of $E_\text{drift} = \text{273~V/cm}$, and it takes about 2.3~ms for electrons to drift across the full width of the TPC.
When the ionization electrons arrive at the anode, their charge is measured by the wire planes. The Y plane is oriented vertically; the U and V planes are oriented at $\pm60^\circ$ with respect to the Y plane. Wires in each plane are spaced 3~mm apart. The U and V planes measure an induced signal from the electrons passing nearby while the Y plane serves as the collection plane.
In wire plane images such as those shown in Figure~\ref{fig:event_display}, each column of pixels represents the signal ($z$-axis) on a single wire ($x$-axis) read out as a function of time ($y$-axis).
The wire front-end electronics are also immersed in the liquid argon, which significantly reduces associated noise.

 \subsection{Deep Learning}
 \label{sec:dl}
Deep learning refers to a broad class of machine learning algorithms that use many layers of non-linear processing units, often called ``neurons'', with each layer learning feature representations based on the layer beneath it to reach increasing levels of complexity and abstraction.
MicroBooNE uses a specific type of deep learning algorithms called convolutional neural networks (CNNs).
CNNs have been developed primarily for image analysis tasks, and in this work we apply them to MicroBooNE wire plane images as shown in Figure~\ref{fig:event_display}.
We use them in two distinct tasks:
\begin{itemize}
\item Classification: The CNN is trained on a set of images, each labeled as being a member of particular category. The CNN is then given a new image and is asked to identify which category it belongs to.
\item Semantic segmentation: The CNN is trained on a set of images in which each pixel is labeled as being a member of a particular category. The CNN is then given a new image and is asked to identify which category each pixel in that image belongs to. For example, if the CNN is trained appropriately and is given an image of a person riding a bicycle, it should label pixels in the image representing the person as ``person'' and the pixels representing the bike as ``bicycle''.
\end{itemize}

\section{Low-Energy Excess Analysis}

\subsection{Definition of the Signal}
\label{sec:signal_def}
For the purposes of this analysis, we define signal events to be contained events with a one lepton, one proton (1$\ell$1p) topology.
The neutrino energy range of interest is 200 to 600~MeV. In addition, the lepton is required to have kinetic energy of at least 35~MeV and the proton is required to have kinetic energy of at least 60~MeV. There may be other protons below this energy threshold.

\begin{figure}
\begin{center}
\includegraphics[width=0.85\textwidth]{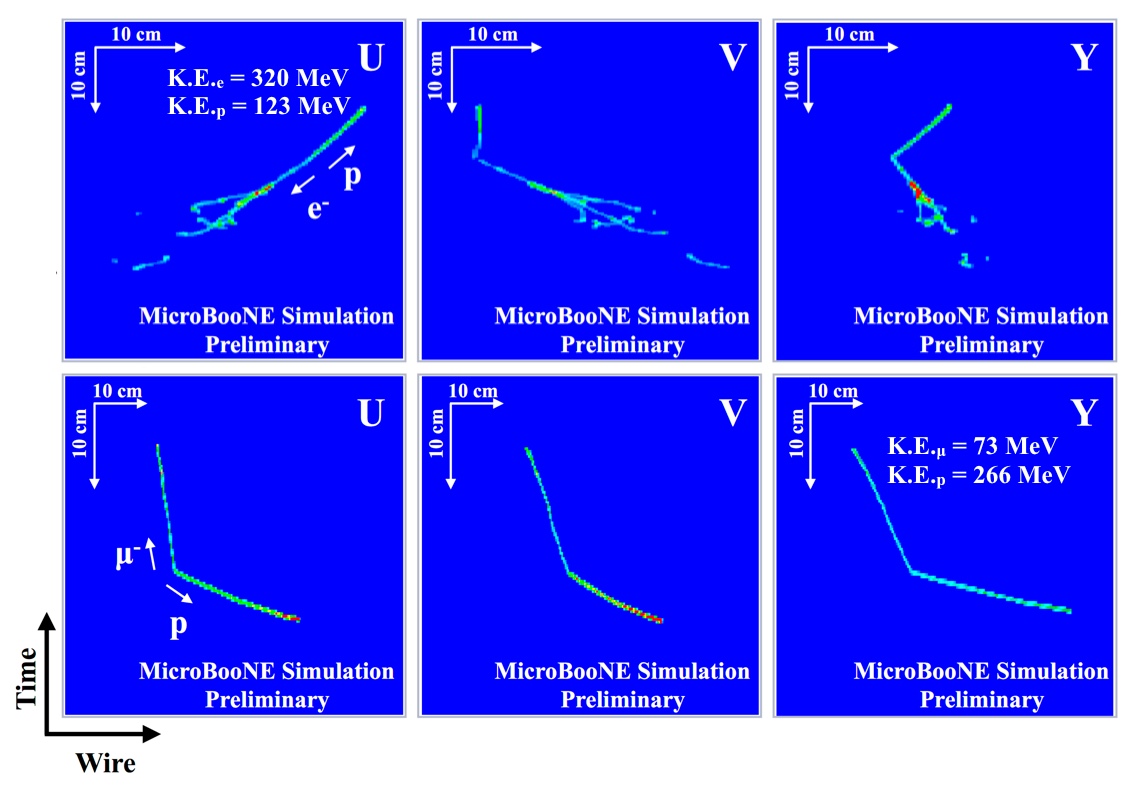}
\caption{Examples of wire plane images produced by simulated neutrino interactions in the MicroBooNE detector. The upper row shows the U, V, and Y wire plane images for a $\nu_e$ event; the lower row shows the same for a $\nu_\mu$ event. Both satisfy the 1$\ell$1p signal criteria described in Section~\ref{sec:signal_def}.}
\label{fig:event_display}
\end{center}
\end{figure}

We expect that imposing these restrictions will allow us to isolate a pure sample of $\nu_e$ events, for which the only significant background is due to intrinsic $\nu_e$ contamination in the BNB beam.
The expected number of intrinsic $\nu_e$ events can be constrained based on a measurement of $\nu_\mu$ events over the same energy range, which puts strong constraints on flux and cross-section systematic uncertainties.
We wish to make a measurement of both $\nu_e$ and $\nu_\mu$ events that meet the 1$\ell$1p criteria described above. An example of each is shown in Figure~\ref{fig:event_display}.

\subsection{Overview of the Reconstruction Chain}
\begin{figure}
\begin{center}
\includegraphics[width=0.5\textwidth]{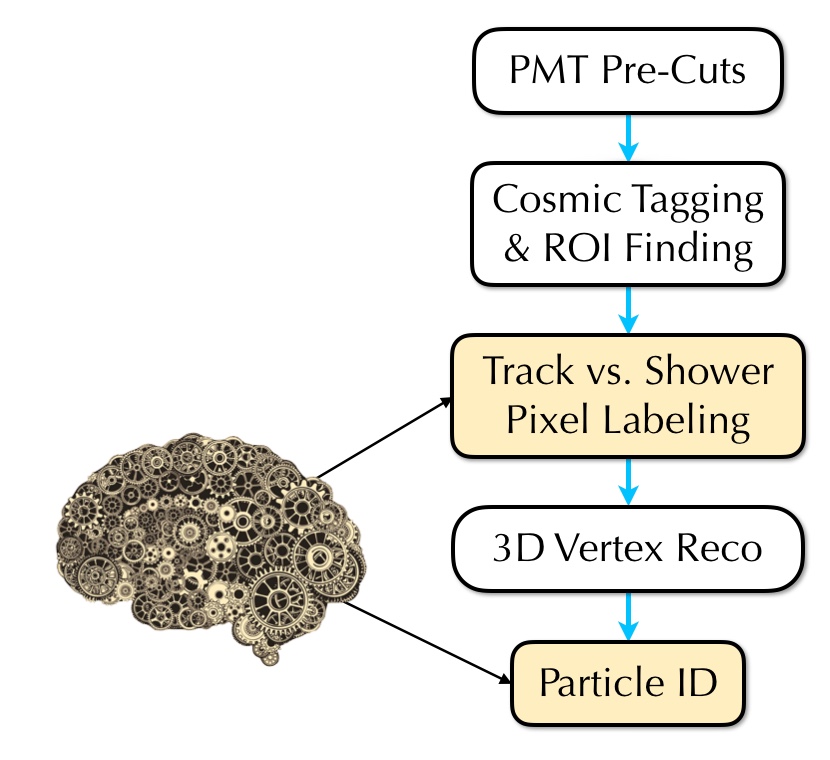}
\caption{An outline of the reconstruction chain described in Section~\ref{sec:reco}. The steps shaded in yellow indicate the use of deep learning. All other steps use traditional algorithms.}
\label{fig:reco_chain}
\end{center}
\end{figure}

The reconstruction chain is outlined in Figure~\ref{fig:reco_chain}. Each of the individual steps is discussed in detail in Section~\ref{sec:reco}.
Note that the reconstruction consists of a combination of traditional and deep learning algorithms. The majority of the steps use traditional analysis techniques. We use CNNs in two steps: for labeling pixels as track-like or shower-like (Section \ref{sec:ssnet}), and for particle identification (Section \ref{sec:pid}). 

\section{Reconstruction Chain}
\label{sec:reco}

\subsection{PMT Pre-Cuts}
We use a set of initial of cuts based on the PMT information to reject low-energy backgrounds.
The cuts have been tuned to maintain more than 96\% of neutrinos, based on their performance on simulated data events. At the same time, they reject more than 75\% of background events, based on their performance on off-beam detector data.
By significantly decreasing the number of events that need to be processed through the rest of the reconstruction chain, the PMT pre-cuts also reduce the computational footprint of our analysis. 

\subsection{Tagging Cosmic Pixels \& Finding Regions of Interest}

The MicroBooNE detector is located on the surface. Given the size of the detector and the relatively long TPC drift time, this results in about 20 cosmic ray muon tracks in every read-out window for the wire planes. These tracks are often several meters long.
In contrast, the low-energy neutrino events that we will measure in this analysis are quite rare and the tracks are of order ten centimeters long.
The goal of this step in the reconstruction chain is to identify and hone in on regions within an event that contain a candidate neutrino interaction.

First, we tag pixels that are associated with cosmic ray tracks.
We use the fact that cosmic tracks will cross the TPC boundary.
The boundary crossing points are identified in one of three ways, depending on the location of the crossing.
	If a track crosses the top or the bottom, it will deposit charge on a unique triplet of U, V, and Y wires that meet there.
	If a track crosses the upstream or downstream face of the TPC, it will deposit charge on the first or last few wires of the Y plane.
	Finally, if a track crosses the anode or cathode, there is a characteristic time difference between when the scintillation light arrives at the PMTs and when the ionization charge arrives at the wires. For an anode crossing, the time difference is negligible. For a cathode crossing, it is the time it takes for electrons to drift across the entire width of the TPC, or about 2.3~ms.
Once the TPC boundary crossing points have been identified, the algorithm follows the associated tracks into the bulk of the detector using a 3D path-finding algorithm to follow lines of deposited charge.
All pixels associated with tracks that cross the TPC boundary are tagged.

Next, the algorithm searches for clusters of untagged pixels, which indicate deposited charge that was contained within the TPC volume. We draw a 3D region of interest (ROI) box around any such clusters. These represent candidate neutrino interactions.

\subsection{Labeling Track and Shower Pixels}
\label{sec:ssnet}
The labeling of track and shower pixels is the first application of deep learning in the reconstruction chain. The goal of this step is to separate tracks and showers, which makes vertex reconstruction and clustering more efficient.

\begin{figure}
\begin{center}
\includegraphics[width=\textwidth]{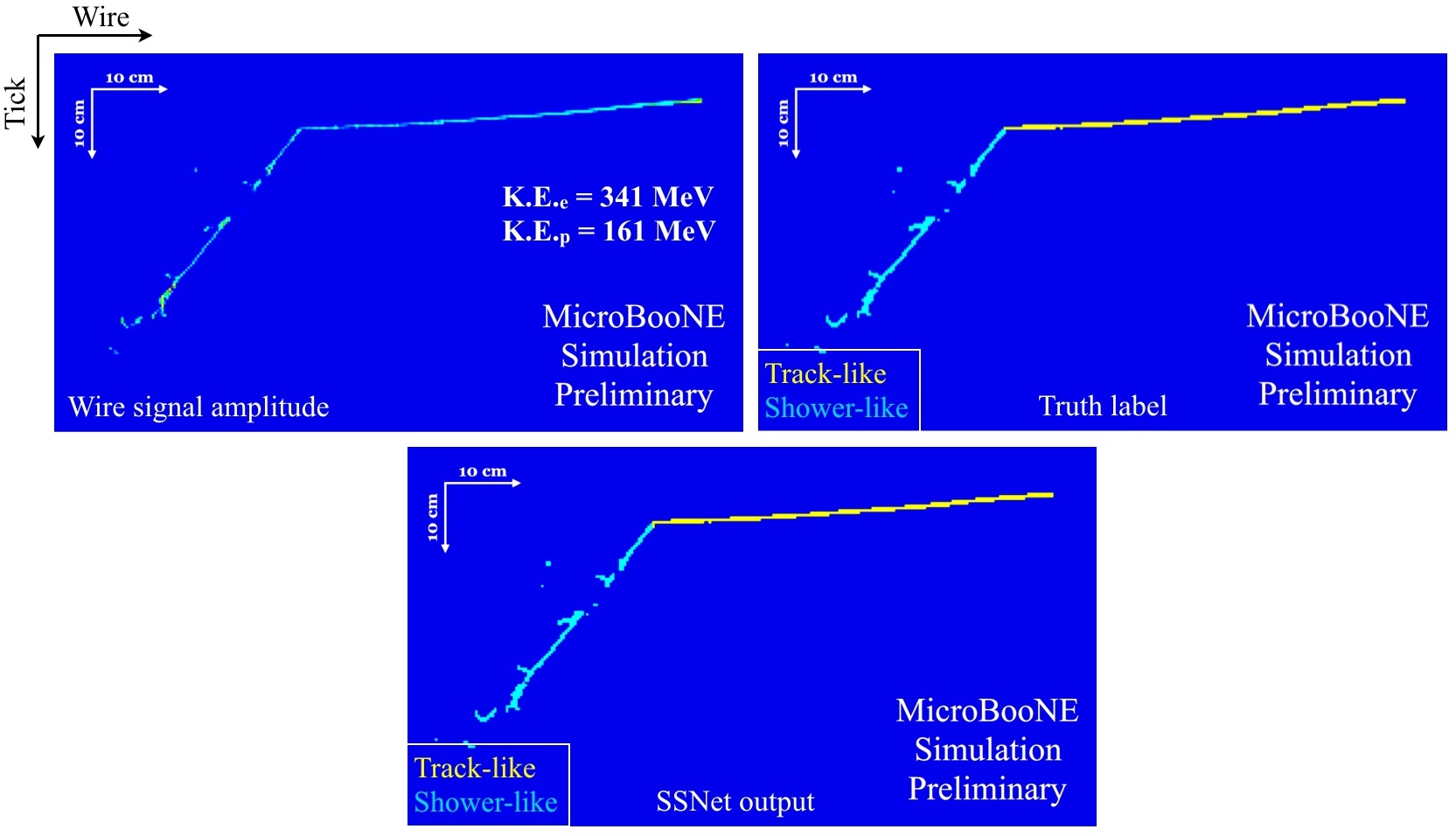}
\caption{Example of the performance of SSNet on a simulated $\nu_e$ event. The upper left panel shows the wire signal amplitude, which is the input to SSNet. The upper right panel shows the simulation-based truth labels for each of the pixels. Background pixels are dark blue, track pixels are yellow, and shower pixels are cyan. The bottom panel shows the output of SSNet using the same color scheme. SSNet labels the pixels with high accuracy.}
\label{fig:ssnet_sim}
\end{center}
\end{figure}

We use a specialized CNN called a semantic segmentation network (SSNet).
As described in Section~\ref{sec:dl}, semantic segmentation labels each pixel in an image as belonging to one of several possible categories.
In this case, we have trained the network to label pixels in MicroBooNE wire plane images as ``background'', ``track-like'', or ``shower-like''.
The output of SSNet for a simulated signal $\nu_e$ event is shown in Figure~\ref{fig:ssnet_sim}.
The performance of SSNet for this task is promising.

\begin{figure}
\begin{center}
\includegraphics[width=\textwidth]{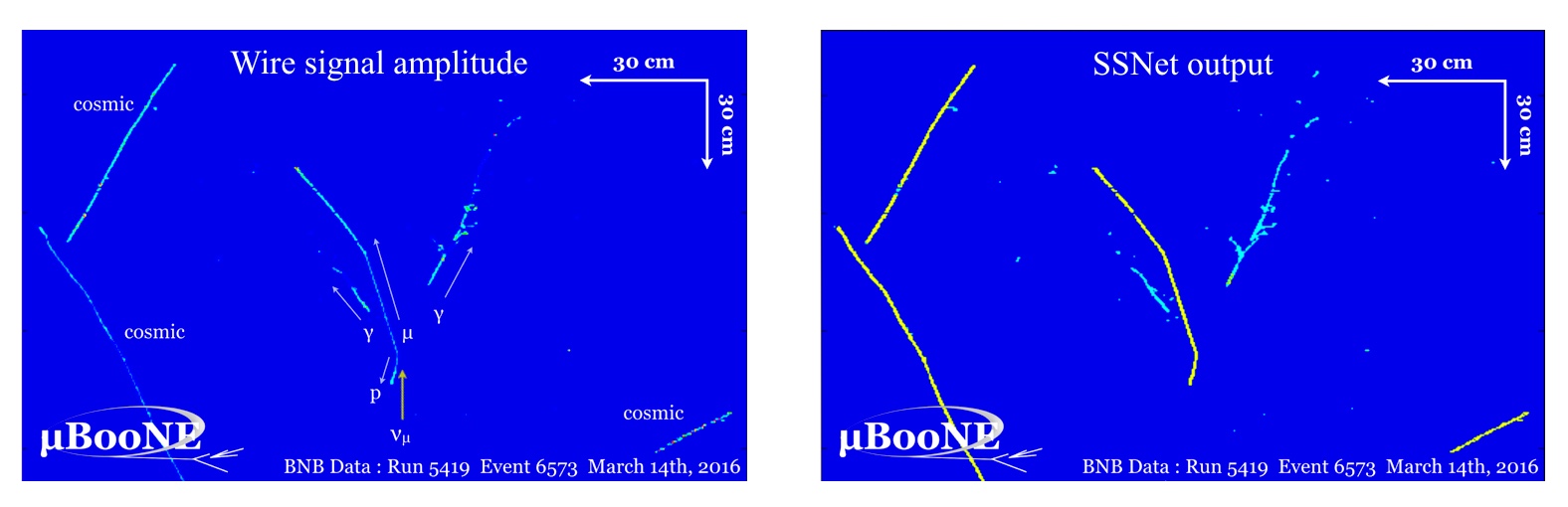}
\caption{Example of the performance of SSNet on a $\pi^0$ event from detector data. The left panel shows the wire signal amplitude with labels indicating the different particles in the event. The right panel shows the output of SSNet using the same color scheme as in Figure~\ref{fig:ssnet_sim}. The muon and proton are correctly labeled as track-like. The photon-initiated electromagnetic showers are correctly labeled as shower-like, except for a small portion of the trunk of one shower that is labeled track-like.}
\label{fig:ssnet_pi0}
\end{center}
\end{figure}

We have also studied the performance of SSNet on detector data.
In particular, we ran SSNet over a sample of selected charged current $\pi^0$ events. The output of SSNet for one of these events is shown in Figure~\ref{fig:ssnet_pi0}.
Compared to a human expert's manual pixel labeling of detector data events, SSNet agrees more than 90\% of the time.
This is similar to the performance on simulated events. 

\subsection{Vertex Reconstruction}
The next step is 3D vertex reconstruction. This proceeds using one of the two methods described below, depending on the results of the previous step.

If the ROI contains both track- and shower-labeled pixels, we are able to use the boundary where the track and shower meet to identify possible vertex points.
In each of the three wire planes, the algorithm finds the endpoint of the cluster of track-labeled pixels where shower-labeled pixels are attached.
These endpoints are then correlated across the wire planes to identify a candidate 3D vertex.
To refine this initial estimate, the algorithm scans through a 3D region about the candidate vertex. At each point in the scan, it calculates how well the hypothesis that the track and the shower come out from the point explains the observed charge deposition across all three planes.
The algorithm places a vertex at the best point out of the region scanned.

If the ROI instead contains only track-identified pixels, we must use a different method to identify possible vertex points.
In each of the three wire planes, the algorithm searches for any kink points in the tracks and creates a 2D vertex seed at each.
Using a scanning method similar to before, we refine the initial estimate of the vertex seed position, although this time treating each plane separately and thus in only two dimensions.
The information from all three planes is then combined.
If the best 2D vertex candidate from each of the three planes are located such that they are consistent with a single 3D point, then the algorithm places a vertex at that point.

After vertex reconstruction has been completed, regions of charge going out from the vertex in distinct directions are clustered together.

\subsection{Particle Identification}
\label{sec:pid}

These clusters are used as the input for a CNN that has been trained to do single-particle classification. We have trained this network to recognize electrons, photons, muons, charged pions, and protons. It returns the particle category that best matches the cluster provided.
The training and performance of this CNN has been described previously~\cite{dl_paper}, and the result is summarized in Table~\ref{tab:pid}.

\begin{table}[b]
\begin{center}
\begin{tabular}{ll} \toprule
Particle & Correct ID \\ \midrule
$e^-$ & $77.8 \pm 0.7 \%$ \\
$\gamma$ & $83.4 \pm 0.6 \%$ \\
$\mu^-$ & $89.7 \pm 0.5 \%$ \\
$\pi^-$ & $71.0 \pm 0.7 \%$ \\
$p$ & $91.2 \pm 0.5 \%$ \\
\bottomrule
\end{tabular}
\caption{Particle classification performance on simulated single-particle images for five particle types using HiRes GoogLeNet architecture~\cite{dl_paper}. Quoted uncertainties are purely statistical and assume a binomial distribution. }
\label{tab:pid}
\end{center}
\end{table}

We expect to improve the performance of this step in the reconstruction chain further.
This single-cluster, single-particle classification does not take into account the larger context of the event, including factors like event topology and vertex separation, which will improve our ability to distinguish different particles.

The main misidentification of an electron is a photon, and vice versa~\cite{dl_paper}. This is to be expected, as both have a shower-like appearance that is distinct from the other particle categories and the CNN is learning to differentiate based on such visual qualities.
However, the ability to correctly classify electrons vs.\ photons about 80\% of the time still represents a significant improvement over MiniBooNE.

\section{Addressing Systematic Uncertainties}
As this analysis progresses, we are also thinking about how to address systematic uncertainties.
This will require a variety of approaches, including some tailored to assess the systematics associated with deep learning techniques. This section describes one such approach.

\subsection{Topological Sidebands}
In general, a sideband is a set of events that are outside the set of events that are the target of the analysis, but have important similarities to them.
A typical sideband sample would be events that are adjacent in kinematic space.
We instead consider events that are similar in topology, or a topological sideband. We believe that these samples will be helpful in assessing systematic uncertainties of deep learning algorithms because they are based on the use of visual information.

In particular, we are interested in drawing topological sidebands from detector data.
We know there are discrepancies between our simulations and detector data, and we have found that CNNs can be sensitive to these differences.
We plan to use data-based topological sideband samples to test simulation versus data agreement, and also to study efficiencies.
We plan to study several different topological sideband samples.

One is charged current $\pi^0$ events, such as the one shown in Figure~\ref{fig:ssnet_pi0}.
These events have a 1$\mu$1p vertex, the same as $\nu_\mu$ events that we will measure.
They can be tagged by the nearby electromagnetic showers resulting from the $\pi^0\to\gamma\gamma$ decay.
Such events have already been used to test the performance of SSNet, and we will also use them for additional studies.

Another sample of interest is neutral current $\pi^0$ events in which one of the photons from pion decay converts into an electromagnetic shower close to the vertex.
This creates an apparent vertex with one shower and one proton coming out, similar to $\nu_e$ events that are our signal, such as the one shown in the upper row of Figure~\ref{fig:event_display}.
They can be tagged by the second electromagnetic shower nearby.

A third sample is cosmic ray muons that stop within the TPC and decay to a Michel electron~\cite{michel_paper}.
The Michel decay vertex has an incoming muon track and an outgoing electromagnetic shower. This vertex has a track and shower topology similar to $\nu_e$ events.
A stopping muon sample also has the advantage of statistics. The number of stopping muons in the MicroBooNE data set is much larger than any type of neutrino interaction.

The final topological sideband sample I will discuss is chimera events, detailed in the following section.

\subsection{Chimera Events}

\begin{figure}
\begin{center}
\includegraphics[width=\textwidth]{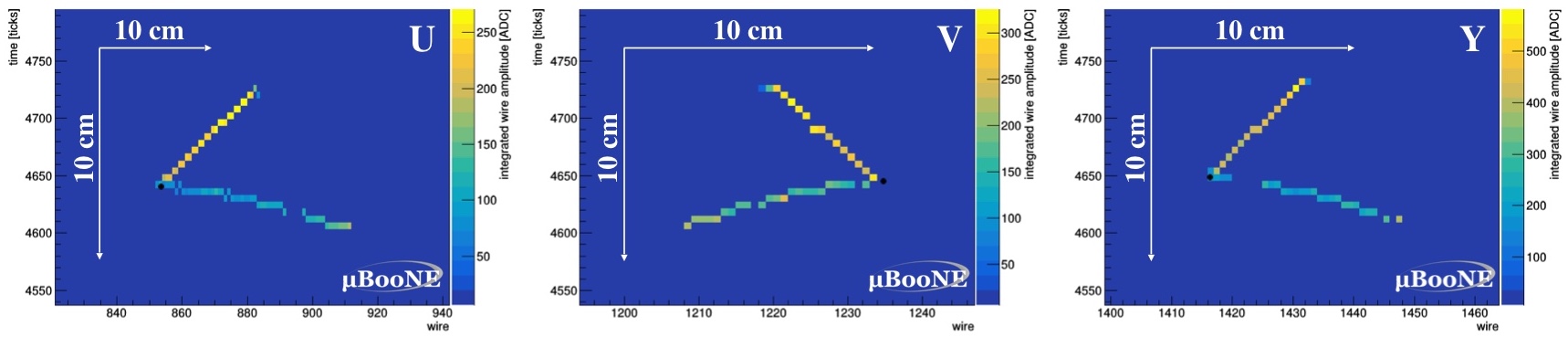}
\caption{Example of an early $\nu_\mu$-like chimera event. The three plots show the hit intensities in the U, V, and Y wire planes. The black dot indicates the location of the chimera vertex.}
\label{fig:chimera}
\end{center}
\end{figure}

Chimera events are created by isolating single-particle components from cosmic ray data, and combining them to create events with a neutrino-like topology. An example of a $\nu_\mu$-like chimera event is shown in Figure~\ref{fig:chimera}.

For $\nu_\mu$-like chimera events, we use a proton and a stopping muon. The entering portion of the muon track is removed so that it appears to be contained in the fiducial volume of the detector.
For $\nu_e$-like chimera events, we would use a proton and an electron or other electromagnetic shower of appropriate energy. These are more difficult to create than the $\nu_\mu$ because of the relative scarcity of the single-particle components.

When combining single-particle components to create chimera events, we allow, for but try to minimize, spatial translations because of slight variations in detector response. Minimizing translations will minimize any differences in detector response corresponding to the two single-particle components being used.
We do not rotate the tracks to avoid modifying the wire-to-wire correlation within each track.

The value of chimera events is the ability to create a data-driven sample that will cover the entire physics parameter space of interest for our signal. These will serve as a useful complement to other data-based topological sideband samples.

\section{Summary}

We have developed a fully automated reconstruction chain for low-energy neutrino events in the MicroBooNE detector. This chain uses both traditional and deep learning algorithms.
The chain is able to: reject cosmic backgrounds, find candidate neutrino interactions within events, separate tracks and showers, reconstruct vertices in 3D, cluster charge outgoing from reconstructed vertices, and identify individual particles.
Deep learning is used for semantic segmentation in track versus shower labeling, and for classification in particle identification.

Work on full 3D reconstruction, including reconstruction of energy deposition along the particle trajectory ($dE/dx$) and final event selection, are ongoing.
Studies of efficiencies and systematic uncertainties are also in progress.

We believe this work incorporating deep learning techniques into a MicroBooNE analysis represents an important development for current and future LArTPC programs.

\subsection*{Acknowledgements}
The author acknowledges support from the National Science Foundation grant NSF-PHY-1505855. On behalf of the MicroBooNE collaboration, we also acknowledge support from the Department of Energy Office of Science.

\end{document}

%% file: econfmacros.tex
\def\beq{\begin{equation}}
\def\eeq#1{\label{#1}\end{equation}}
\def\eeqn{\end{equation}}

\def\beqa{\begin{eqnarray}}
\def\eeqa#1{\label{#1}\end{eqnarray}}
\def\eeqan{\end{eqnarray}}

\let\bar=\overbar

\def\Dslash{\not{\hbox{\kern-4pt $D$}}}
\def\dslash{\not{\hbox{\kern-2pt $\del$}}}

\def\msb{{\bar{\ssstyle M \kern -1pt S}}}

